\documentclass[11pt,aps,a4paper,superscriptaddress,twocolumn]{revtex4-2}
\usepackage
[paperwidth=22cm,paperheight=29.7cm,left=1.5cm,right=1.5cm,
top=1.5cm,bottom=2.2cm]
{geometry}
\usepackage{graphicx,subfigure}
\usepackage{bm}
\usepackage{array}
\usepackage{slashed}
\usepackage{braket}
\usepackage{multirow}
\usepackage{bigstrut}
\usepackage[small]{caption}
\usepackage{hyperref}
\usepackage{amsmath}
\usepackage{amssymb}
\usepackage{wasysym}
\usepackage{latexsym}
\usepackage{stmaryrd}
\usepackage{txfonts}
\usepackage[T1]{fontenc}
\usepackage{textcomp}
\usepackage{xcolor}
\usepackage{titlesec}
\usepackage{comment}
\usepackage{color}
\usepackage{float}
\usepackage{setspace}

\usepackage[dvipsnames]{xcolor}

\usepackage{float}
\usepackage{amsmath,amsfonts,amssymb,epsfig,graphicx}
\usepackage[normalem]{ulem}

\hypersetup{
	colorlinks=true,       % false: boxed links; true: colored links
	linkcolor=blue,        % color of internal links
	citecolor=blue,        % color of links to bibliography
	filecolor=magenta,     % color of file links
	urlcolor=blue         
}

\definecolor{lightgray}{gray}{0.85}
\definecolor{darkblue}{rgb}{0.0, 0.0, 0.55}
\definecolor{darkred}{rgb}{0.7, 0.0, 0.0}
\definecolor{darkgreen}{rgb}{0.0, 0.59, 0.0}

\newcommand{\qbar}{\bar{q}}
\newcommand{\ubar}{\bar{u}}
\newcommand{\Ubar}{\bar{U}}
\newcommand{\dbar}{\bar{d}}
\newcommand{\Dbar}{\bar{D}}
\newcommand{\sbar}{\bar{s}}
\newcommand{\cbar}{\bar{c}}

\begin{document}

\setstretch{1.}
\newcommand*{\hpdf}{HERAPDF2.0}

\newcommand{\XJU}{School of Physical Science and Technology, Xinjiang University, Urumqi 830017, China}\affiliation{\XJU}

\newcommand{\PKU}{School of Physics, Peking University, Beijing 100871,
China}\affiliation{\PKU}

\newcommand*{\ZZU}{School of Physics, Zhengzhou University, Zhengzhou 450001, China}\affiliation{\ZZU}

\author{Alim~Ruzi}\email{alim.ruzi@pku.edu.cn}
\affiliation{\XJU}\affiliation{\PKU}
\author{Bo-Qiang Ma}\email{mabq@pku.edu.cn}
\affiliation{\ZZU}\affiliation{\PKU}%\affiliation{\CHEP}

\title{Redetermination of proton sea distributions}

\begin{abstract}
The shapes of light flavor sea quark distributions of the proton are examined directly from two rounds of NNLO global analysis of HERA $e^{\pm}p$ deep inelastic scattering cross section measurements, termed as HERAshape and the ATLAS measurement of $W^{\pm}/Z$ production from $pp$ collision at $\sqrt{s}$ = 7 TeV, termed as ATLASshape. An asymmetric distribution between anti-up ($\ubar$) quark and anti-down ($\dbar$) quark is found in both analysis, showing that the anti-up quark distribution exceeds over anti-down quark distribution in the momentum fraction range $x\in(10^{-2}, 1)$ of these partons. The Gottfried Sum Rule is reevaluated from these extracted parton distribution functions and the obtained value differs surprisingly from that of the NMC and the NuSea Collaborations.
\end{abstract}


%\date{\today}
\maketitle

It is well known that nucleons are constructed from point-like particles, denominated as partons, revealed by  lepton nucleon Deep Inelastic Scattering (DIS) experiment~\cite{Bloom:1969kc,Breidenbach:1969kd} for decades ago. The internal structure of nucleons, quantified by Parton Distribution Functions (PDFs), is well determined up to a very high precision in the valence quark and gluon content~\citep{DelDebbio:2007ee, Dulat:2015mca, Harland-Lang:2014zoa, CooperSarkar:2011aa}. However, the sea quark distributions are still ambiguous in our current understanding of internal structure of nucleons, especially the anti-up ($\ubar$) and anti-down ($\dbar$) quark distributions. The improved result of asymmetric distributions of the $\ubar$ and $\dbar$ quarks was first found in the violation of Gottfried Sum Rule (GSR)~\cite{Gottfried:1967kk}, reported by the New Muon Collaboration (NMC)~\cite{Amaudruz:1991at} from deep-inelastic muon scattering on hydrogen and deuterium targets. Measurements from muon pair production through the Drell-Yan (DY) process in proton scattering on hydrogen and deuterium targets by the NA51 experiment~\cite{Baldit:1994jk} and pion production from semi-inclusive DIS of positron on hydrogen and deuterium targets by the HERMS collaboration~\cite{Ackerstaff:1998sr} also showed a large asymmetry in the distributions of up and down anti-quarks in nucleon. Another analysis of improved measurements of $\dbar / \ubar $ and $\dbar -\ubar $ from the cross section ratio of proton hydrogen scattering to proton deuteron scattering by the NuSea collaboration~\cite{Hawker:1998ty,Towell:2001nh}  also confirmed the asymmetric distributions of $\dbar $ and $\ubar $ of the proton, proposing there is an excess of $\dbar$ over $\ubar$. However, a number of factors should be considered carefully because of the usage of nuclear targets in these experiments. These factors include target mass effects, higher-order corrections and nuclear shadowing of the virtual photon in the deuteron~\cite{Georgi:1976ve,Badelek:1994qg}. The target-mass effect could be ignored while the higher order corrections and nuclear shadowing effect, to some extent, have impacts on the GSR.

The Gottfried sum, $S_G$, simply expressed as $\int_0^1[F_2^p(x)-F_2^n(x)]dx/x$, casts into the following form if expressed in terms of parton distributions
%\begin{equation}
\begin{align}
\begin{split}
    S_G &=  \int_0^1 \sum_i Q^2_i\left[q_i^p(x) + \qbar _i^p(x) - q_i^n(x) - \qbar _i^n(x)\right] dx \\
    &=\frac13+\frac29 \int_0^1 \left[ 4 \big( \ubar^p(x) - \ubar^n(x)\big)+\big(\dbar^p(x) - \dbar^n(x)\big) \right]dx, 
\label{eq:gs}
\end{split}
\end{align}
%\end{equation}
where $Q_i$ corresponds to the charge of a quark of flavor $i$, $q_i(x)^p$ and $q_i^n(x)$ are the quark distributions of the proton and the neutron. Under the assumption of isospin symmetry between the proton and the neutron, i.e., $\ubar^p = \dbar^n = \ubar$ and  $\dbar^p = \ubar^n = \dbar $, Eq.~(\ref{eq:gs}) reduces to 
%\begin{align}
\begin{equation}
S_G = \frac13 +\frac23 \int_0^1\left[\ubar(x) - \dbar(x) \right ]dx.
\label{eq:gsQPM}
\end{equation}
%\end{align}
If the sea SU(2) symmetry is further assumed, $\ubar(x) = \dbar(x)$, we obtain the Quark Parton Model (QPM) prediction for GSR, to be 1/3. The NMC result, $0.240 \pm 0.016$~\cite{Amaudruz:1991at}, is roughly $10\%$ lower than this value. 
This discrepancy between the NMC result and QPM prediction of the GSR was considered as an indication of strong violation of this sum rule, from which the idea that there is an excess of $\dbar$ over $\ubar$ inside proton can be infered according to  Eq.~(\ref{eq:gsQPM}). This result requires a good isospin symmetry between proton and neutron to hold exactly among the light flavor sea distributions. However, in Refs.~\cite{Ma:1991ac, Ma:1992gp, Ma:1994pe}, it is argued  that the violation of GSR could also be due to the isospin symmetry breaking while preserving the symmetric distribution in the sea of up and down quarks of the proton, or the combined effect of both the isospin symmetry breaking among nucleons and an asymmetric distribution in $\ubar$ and $\dbar$. 

 If we assume that flavor symmetry in the sea of nucleon holds exactly, $\ubar ^p = \dbar ^p = \qbar ^p$, $\ubar^n =  \dbar ^n = \qbar ^n$, then Eq.~(\ref{eq:gs}) reads as
%\begin{align}
\begin{equation}
S_G = \frac 13 + \frac{10}{9}\int_0^1\left[\qbar ^p(x) - \qbar ^n(x)\right]dx.
\label{eq:gsr3}
%\end{align}
\end{equation}

Thus, we can interpret the discrepancy between the NMC and QPM predictions, according to Eq.~(\ref{eq:gsr3}),  as an indication that there are more sea quarks in the neutron than in the proton. The contribution from isospin symmetry breaking to the GSR is expected to be small~\cite{Gottfried:1967kk}, but this symmetry does not predict equality of $\ubar $ and $\dbar $ distribution neither. So there is an ambiguity in the light flavor distribution of the proton sea. This motivates us to reconsider $\ubar $ and $\dbar $ shapes of the proton. 

Taking into account the limitations of the nuclear fixed target experiment mentioned earlier, we suppose a direct examination on the shapes of light flavor sea distributions of the proton, especially on $\ubar$ and $\dbar$, from a pure proton-related experiment, which is sensitive to the flavor decomposition of the proton, be necessary to tackle the current issue. Luckily, the HERA H1 and ZEUS combined measurements of $e^{\pm}p$ inclusive DIS experiment~\cite{Abramowicz:2015mha} and the gauge boson production in proton-proton collision experiment at ATLAS~\cite{Aad:2011dm, Aaboud:2016btc}  provide us such ideal data sets. In the present letter, we carry out a NNLO pQCD analysis to extract the light flavor sea quarks distribution of the proton, firstly, using combined HERA H1 and ZEUS $e^{\pm}p$ data, secondly, using both HERA and ATLAS measurements of $W^{\pm}/Z$ production cross sections. Meanwhile, the $S_G$ is reevaluated using the extracted $\ubar$ and $\dbar $ quark distributions.

The combined $e^{\pm}p$ cross section data can cover a wide range in two kinematic variables: $0.045 \leq Q^2 \le 50000$ GeV$^2$, $6 \times 10^{-7} \leq x \leq 0.65$ in neutral current (NC) interaction (mediated by $\gamma^*/Z$) and $ 200 \le Q^2 \le 50000$ GeV$^2$, $1.3 \times 10^{-2} \le x \le 0.40$ for charged current (CC) interaction (mediated by $W^{\pm}$), with electron beam energy of $E_e \simeq 27.5$ GeV and proton beam energies of $E_p = $ 920, 820, 575, 460 GeV. The obtained PDFs from these data are free of any hadronic corrections and the assumption of isospin symmetry in the nucleon. Another important feature of the HERA data is that the subprocess $W^+~\{d,~s\}\to \{u,~c\}$ and $W^-~\{\dbar,~\sbar\}\to \{\ubar,~\cbar\}$ in CC interaction has a direct sensitivity to the light sea flavor separation, especially to $\ubar$ and $\dbar$.
 
The HERA $e^{\pm}p$ DIS data has been the central data set in the determination of PDFs among worldwide PDF working groups. The HERA collaboration uses these data in their \hpdf ~analysis~\cite{Abramowicz:2015mha} with the following parameterization form at the initial transferred four momentum $Q^2_0$ = 1.9 GeV$^2$
\begin{subequations}
\begin{align}
	xg(x)   & = A_gx^{B_g}(1-x)^{C_g}-A'_gx^{B'_g}(1-x)^{C'_g}, \label{eq:xg}  \\
	xu_v(x) & = A_{u_v}x^{B_{u_v}}(1-x)^{C_{u_v}}(1+E_{u_v}x^2), \label{xq:xuv}\\
	xd_v(x) & = A_{d_v}x^{B_{d_v}}(1-x)^{C_{d_v}}, \label{eq:xdv}\\
	x\Ubar (x) &= x\ubar(x)  = A_{\Ubar}x^{B_{\Ubar}}(1-x)^{C_{\Ubar}}(1+D_{\Ubar}x), \label{eq:xU}\\
	x\Dbar(x) &= x\dbar(x)+x\sbar(x) = A_{\Dbar}x^{B_{\Dbar}}(1-x)^{C_{\Dbar }}, \label{eq:xD} \\
	xs(x)& = x\sbar(x) = xf_s\Dbar(x), \label{eq:xs}
\end{align}
\label{eq:heraparam}
\end{subequations}
where $f_s$ is the strange quark fraction factor and by~(\ref{eq:xD}),~(\ref{eq:xs}), we get the relation between anti-down and strange quark as
\begin{equation}
\frac{\sbar (x)}{\dbar(x)} = \frac{f_s}{1-f_s}.
\label{eq:sbdb}
\end{equation}
~~The value of $f_s$ is chosen to be 0.4 as the compromise between the determination of suppressed strange sea from neutrino-induced di-muon production~\cite{Abdallah:2009aa, Nadolsky:2008zw} and another determination of unsuppressed strange sea distribution published in Ref.~\cite{Aad:2012sb} by the ATLAS collaboration. The normalization parameter $A$ is constrained by quark-number sum rule and momentum sum rule. The $B$ parameters, $B_{\Ubar}$ and $B_{\Dbar}$ are set equal in the \hpdf~. The restriction $A_{\Ubar}$ = $f_s(1-A_{\Dbar})$ is implemented on $A$ of $\Ubar$ and $\Dbar$ by hand. These settings forcibly make $x\ubar$ = $x\dbar$ as $x \to0$. Actually, there is no any theoretical proof for these restrictions to be made. The strangeness factor, $f_s$, on which the determination of strange quark is dependent, is given an arbitrary value. This factor, in turn, affects indirectly the exact distribution of $\dbar$ in the proton. 

Considering the above issues, in our first round of fit, we use the HERA $e^{\pm}p$ data and apply the parameterization form in Eq.~(\ref{eq:heraparam}) with freeing the parameters $A,~B$ of $x\Dbar$ and $x\Ubar$. As for the strange quark factor, $f_s$, it is reported that the actual distribution of the strange quark is not suppressed in the proton~\cite{Aaboud:2016btc, Aad:2012sb}. The ratio of $\sbar/\dbar$ is measured  to be $1.19^{+0.15}_{-0.16}$ with great precision by ATLAS collaboration, using the HERA $e^{\pm}p$ data jointly with $W^{\pm}/Z$ production cross section data~\cite{Aaboud:2016btc}. By Eq.~(\ref{eq:sbdb}), we have $f_s$ = 0.54 in our analysis.
%%%%%%%%%%%%%%%%%%%%%%%%%%%%%%%%%%%%%%%%%%%%%%%%%%%%%%%%%%%%%%%%%%%%%%%%%%%%%%%%%
%%%%%%%%%%%%%%%%%%%%%%%%%%%%%%%%%%%%%%%%%%
 \begin{figure*}[bht!]
  \begin{center}
	\includegraphics[scale=0.65]{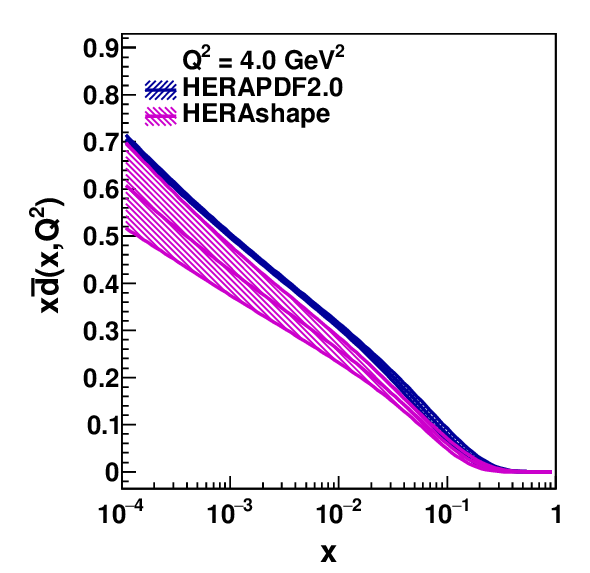}  
	\includegraphics[scale=0.65]{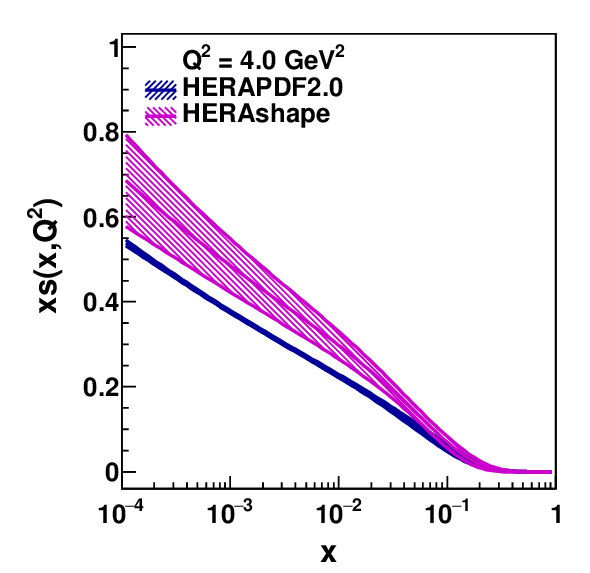}
     \caption{The parton distribution functions $x\dbar(x)$ and $xs(x)$ of HERAshape at $Q^2$ = 4 GeV$^2$ are compared to those of HERAPDF2.0.}
   \label{fig:heraPDF}
 \end{center}
\end{figure*}

By relaxing the restricted parameters in Eq.~(\ref{eq:heraparam}), the number of free fit parameters then adds up to 16 from the 14 fit parameters of  \hpdf. The fitting procedure is done using \texttt{xFitter}~\citep{Alekhin:2014irh,xfitter} open source framework. The theoretical value for $e^{\pm}p$ scattering cross section is calculated by convoluting the parton distribution functions, evolved from DGLAP~\cite{Gribov:1972ri, Gribov:1972rt, Dokshitzer:1977sg, Altarelli:1977zs} evolution equations to the scale of measurements, with the hard scattering cross sections of electron parton scattering, obtained via QCDNUM~\cite{Botje:2010ay} program. For the minimization of $\chi^2$ function, the CERN MINUIT~\cite{James:1975dr} program, interfaced with \texttt{xFitter}, is used. The experimental uncertainties in the HERA $e^{\pm}p$ and the ATLAS $W^{\pm}/Z$ data are contributed to the extracted PDFs by Hessain Error method~\cite{Pumplin:2001ct}. All the relevant electroweak and QCD parameters and the form of $\chi^2$ function are applied following the analysis set up in Ref.~\cite{Abramowicz:2015mha}. We nominate our first fitted PDFs as HERAshape PDFs. The fit yields value of 1353/1128 the fit criteria, $\chi^2$/dof, which is quite well in agreement with that of the HERAPDF2.0. As a result of using the relaxed parameters and the unsuppressed strangeness factor,  Fig.~\ref{fig:heraPDF} shows that the $x\dbar$ distribution decreases roughly more than 10$\%$ while the strange quark distribution increases significantly. It is easy to understand that the strange and the anti-down quark distributions in the original HERAPDF2.0 are lack of reliability. As stated earlier, the HERA $e^{\pm}p$ data provides informations on $\ubar$ and $\dbar$ separation, but can not discriminate between down and strange quarks well when the transferred energy scale is between strange and charm quark threshold.

The ATLAS $W^{\pm}/Z$ data include the cross section measurements of $W^{\pm}$ production in the leptonic decay chanel: $W^+\to l^+\nu$,~ $W^-\to l^-\bar{\nu}$ and $\gamma^*/Z \to ll$ ($l=e, \mu$) DY production process at $\sqrt s =$ 7 TeV with an integrated luminosity of $4.6$ fb$^{-1}$. 
 These data, when used together with the HERA $e^{\pm}p$ data, can constrain the  $u,~\ubar,~d,~\dbar,~g$ and $s+\sbar$ distributions at initial scale~\cite{Martin:2009iq, Gao:2017yyd}. 
In our second round of fit, we use the HERA $e^{\pm}p$ and the ATLAS $W/Z$ data together applying the similar parameterization pattern of ATLAS-series PDFs~\citep{Aaboud:2016btc, Aad:2012sb, Cooper-Sarkar:2018ufj} for  $xg(x), ~xu_v(x), ~xd_v(x)$, $x\ubar(x), ~x\dbar(x), ~x\sbar(x) $. The functional form for gluon, valence-up and valence-down quarks are same with Eqs.~(\ref{eq:xg}),~(\ref{xq:xuv}), and~(\ref{eq:xdv}), while for the rest anti-quarks we apply the following form
\vspace{-2.5mm}
\begin{subequations}
    \begin{align}
    x\ubar(x) & = A_{\ubar}x^{B_{\ubar}}(1-x)^{C_{\ubar}}(1+D_{\ubar}x) \label{eq:xbaru} ,\\
    x\dbar (x) & = A_{\dbar }x^{B_{\dbar }}(1-x)^{C_{\dbar }}(1+D_{\dbar}x) \label{eq:xbard} ,\\
    x\sbar(x) & = A_{\sbar}x^{B_{\sbar}}(1-x)^{C_{\sbar }}. \label{eq:xbars}
    \end{align}
    \label{eq:seaparam}
\end{subequations}
\vspace{-0.5mm}

%%%%%%Four PDFs
\begin{figure*}[bht]
  \begin{center}
    \includegraphics[scale=0.65]{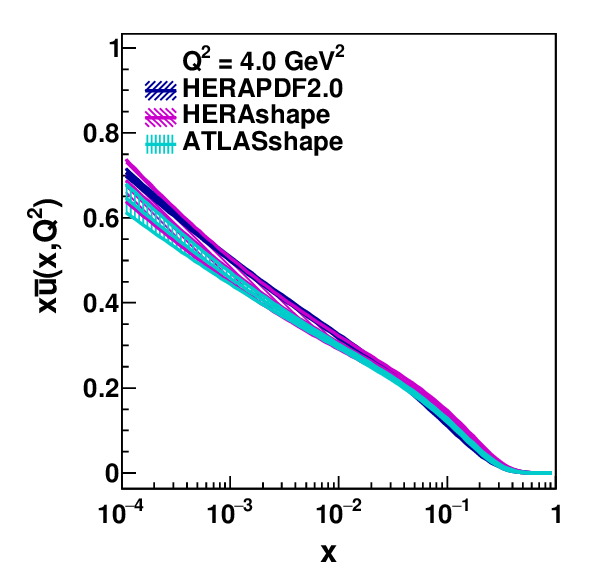}
    \includegraphics[scale=0.65]{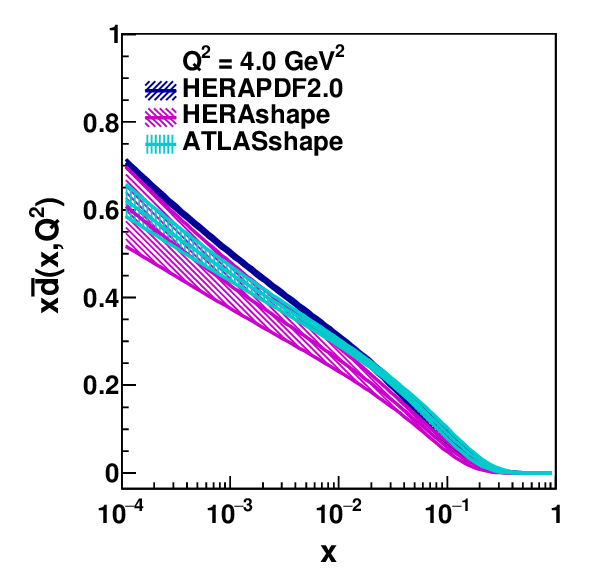}
    \includegraphics[scale=0.65]{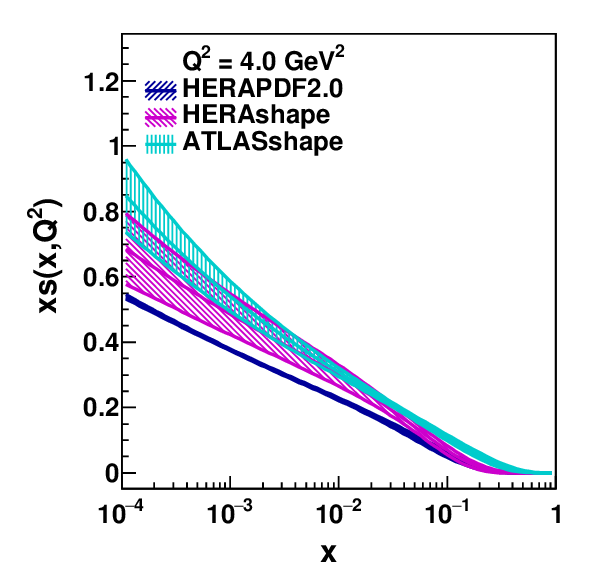}
    \includegraphics[scale=0.65]{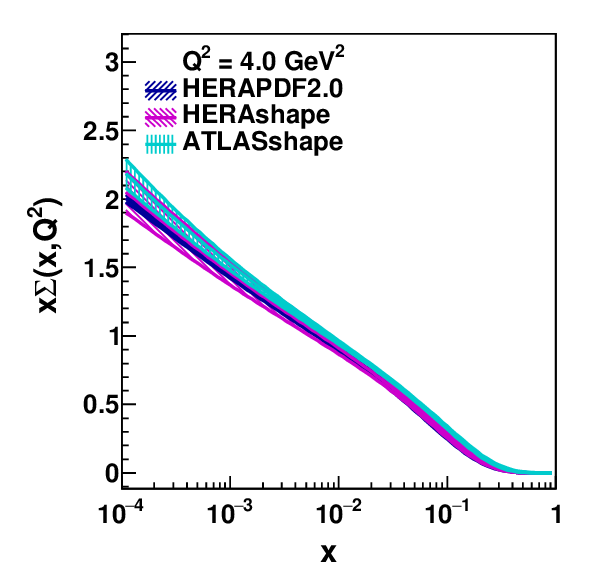}
  \caption{ The parton distribution functions of $x\ubar(x)$, $x\dbar(x)$, $xs(x)$ and $x\Sigma(x)$ of the HERAPDF2.0, HERAshape, and the ATLASshape PDFs at $Q^2$ = 4 GeV$^2$.}
   \label{fig:4pdfs}
 \end{center}
\end{figure*}

%%%Sea difference and ratio
\begin{figure*}[bht!]
  \begin{center}
    \includegraphics[scale = 0.65]{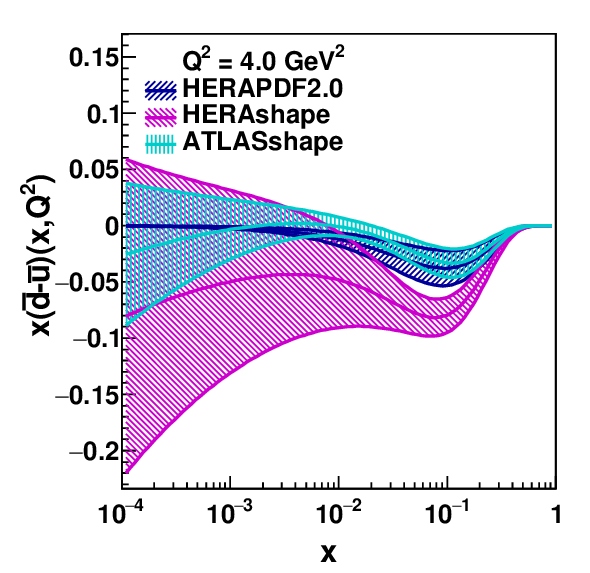}
    \includegraphics[scale = 0.65]{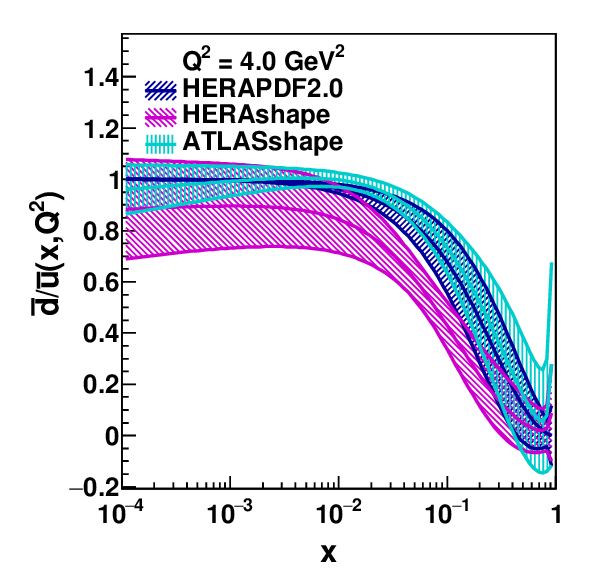}
  \caption{ The sea difference (left) and ratio distribution (right) for HERAPDF2.0, and our analysis: HERAshape, ATLASshape.}
   \label{fig:seadifRatio}
 \end{center}
\end{figure*}  

The differential cross section for $W^{\pm}/Z$ productions are calculated using the APPLGRID program~\cite{Carli:2010rw} interfaced with \texttt{xFitter} at NLO and $K$-factor technique to correct from NLO to NNLO predictions. 
We designate the PDF obtained from our second fit as ATLASshape PDF. A value of 1472/1187 for the total $\chi^2$/dof is obtained. The ATLASshape PDF can describe both HERA and ATLAS data well. The compared PDFs, $x\ubar ,~ x\dbar,~xs,$ and $x\Sigma = 2x(\ubar + \cbar + \dbar + \sbar)$ are shown in Fig.~\ref{fig:4pdfs}. The $x\ubar$, $x\dbar$, $x\sbar$ distributions of the ATLASshape PDFs are  well constrained with smaller uncertainties relative to the HERAshape PDFs. With the inclusion of the ATLAS $W^{\pm}/Z$ data and the application of the parameterization form in Eq.~(\ref{eq:seaparam}), $x\ubar$ is slightly diminished in ATLASshape PDFs with respect to both the HERAPDF2.0 and the HERAshape PDFs. $x\dbar$ keeps almost the same value with HERAshape PDFs but with reduced uncertainty. The strange quark is enhanced visibly, affected by $W^{\pm}/Z$ data. This unsuppressed strange content of proton disagrees the statement of strange suppression in Refs.~\cite{Goncharov:2001qe,  Tzanov:2005kr,Lai:2010vv, Samoylov:2013xoa}, see more detailed discussions in Ref.~\cite{Ruzi:2021znm}. The sea composition in total slightly increases.
%%%%%%%%%%%%%%%%%%%%%%%%%%%%%%%%%%%%%%%%%%%%%%%%%%%%%%%%%%%%%%%%%%%%%%%%%%%%%

 Fig.~\ref{fig:seadifRatio} shows the difference of sea flavors, $x(\dbar - \ubar)$ and its ratio, $\dbar /\ubar$ in three PDF sets. It can be seen that the uncertainty in $x(\dbar- \ubar)$ is reduced significantly after including ATLAS $W^{\pm}/Z$ data relative to HERAshape PDFs, proving that the ATLAS data has good constraints on $\ubar$ and $\dbar$ content of the proton. The bold center line in the figure shows the central value both for $x(\dbar -\ubar)$ and $\dbar / \ubar$. As an update, the ATLASshape PDFs has two zero-cross points in $x(\dbar - \ubar)$ distribution at $ x \approx 10^{-3},~2 \times 10^{-2}$ and below the zero line at other $x$ region. The ratio $\dbar /\ubar$ for all the three PDF sets has similar properties too. So an important conclusion can be made: high energetic proton posses more $\ubar$ quarks than $\dbar$ quarks, or in other words, the densities of anti-down quark is suppressed relative to anti up quark densities.

\begin{figure*}[bht]
  \begin{center}
    \includegraphics[scale = 0.5]{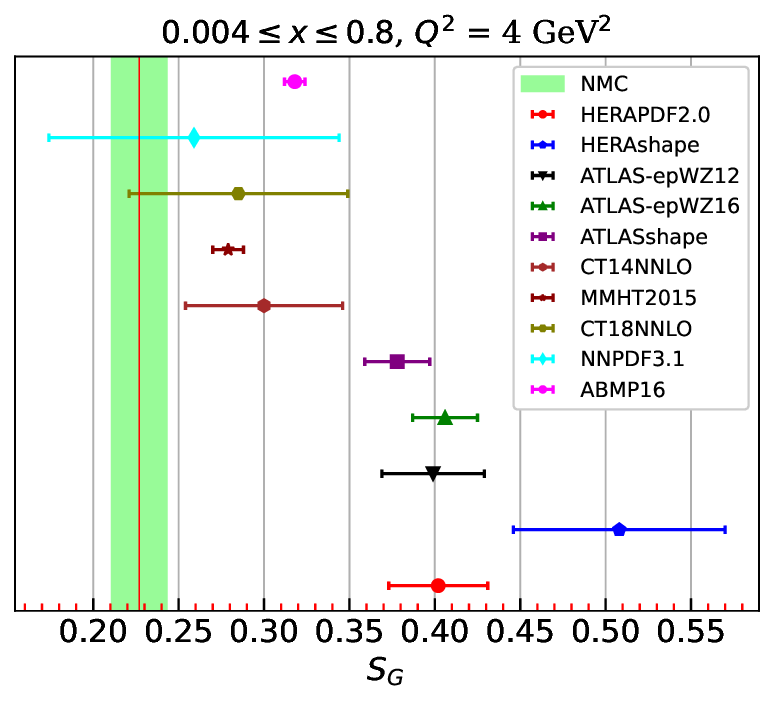}
    \includegraphics[scale = 0.5]{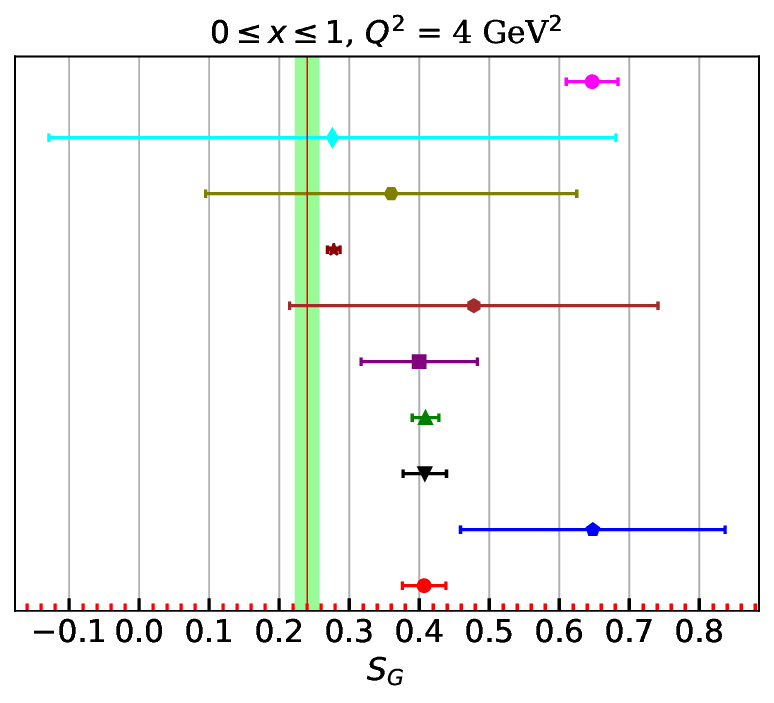}
  \caption{ The computed $S_G$ from HERAPDF2.0, HERAshape, ATLASshape and from other up-to-date worldwide PDFs at $Q^2$ = 4 GeV$^2$.}
   \label{fig:sgpdf}
 \end{center}
\end{figure*}
Finally, we reevaluate $S_G$ in Eq.~\ref{eq:gsQPM} from our analysis and from CT14NNLO~\cite{Dulat:2015mca}, ATLAS-epWZ~\citep{Aaboud:2016btc,Aad:2012sb}, CT18NNLO~\citep{Hou:2019efy}, MMHT2015~\citep{Harland-Lang:2019pla}, NNPDF3.1~\citep{Bertone:2017bme, Manohar:2017eqh}, ABMP16~\citep{Alekhin:2018pai}, using Eq.~(\ref{eq:gsQPM}). The theoretical predictions obtained from these PDF sets are compared with the NMC results within two $x$ ranges: $0.004 \leq x \leq 0.8$ and $0 \leq x \leq 1$ at $Q^2 = $ 4 GeV$^2$. The results are visualized in Fig.~\ref{fig:sgpdf}. The pale green bands in both figure represents the NMC results with $0.227\pm0.0016$ and $0.240\pm0.016$ corresponding in two $x$ ranges.
 
 The $S_G$ obtained from NNPDF3.1, CT18NNLO crosses with that of the NMC within their large uncertainties, with the central values being larger than NMC result in $ 0.004 \leq x \leq 0.8$, shown in the left panel of Fig.~\ref{fig:sgpdf}. The $S_G$ obtained from ABMP16, CT14NNLO, MMHT2015 PDFs agrees with that of NNPDF3.1 and CT18NNLO within their uncertainties but are inconsistent with NMC results.
In the uncertainty range of these PDFs, we are unable to reach to a conclusion on the exact shapes of $x(\dbar - \ubar)$. On the other hand, the HERAPDF2.0, ATLAS-epWZ12, ATLAS-epWZ16 and ATLASshape PDF determinations of the $S_G$ give agreeable results within their uncertainties (the filled-square, down-triangle, triangle and filled circle marks), indicating that there is an excess of $\ubar$ over $\dbar$. The HERAshape PDF gives the largest $S_G$ among all the PDFs, quantifying even larger $\ubar$ content than $\dbar$.  
%%%%%%%%%%%%%%%%%%%%%%%%GSUM figures%%%%%%%%%%%%%%%%%%%%%%%%%%%%%%%%%

  When the $S_G$ is evaluated in whole $x$ range,  $0 \leq x \leq 1$ as shown in the right panel of Fig.~\ref{fig:sgpdf}, the NNPDF3.1, CT18NNLO, CT14NNLO determinations of $S_G$ cross with the corresponding NMC determination within their large uncertainties. Again, because of the large uncertainties of these PDFs, we can not judge the exact asymmetric shape of anti-up and anti-down quarks. However, the $S_G$ from HERAPDF, ATLAS-epWZ12, ATLAS-epWZ16 and ATLASshape PDFs are very consistent with each other, while the HERAshape and ABMP16 PDFs reach to roughly the same value. Therefore, the $S_G$ evaluated from our results, as well as from the up-to-date PDF sets worldwide, confirms that proton has more anti-up quarks than anti-down quarks.
%%%%%%%%%%%%%%%%%%%%%%%%%%%%%%%%%%5

In summary, we examined the exact shapes of anti-up and anti-down quark distribution in the proton and reevaluated the Gottfried sum rule, through a NNLO global analysis of the differential cross section measurements of $e^{\pm} p$ DIS HERA data, denoted as HERAshape PDF and together with the differential ATLAS $W^{\pm}/Z$ production cross sections in $pp$ collision at $\sqrt s$ = 7 TeV, named as ATLASshape PDF, within \texttt{xFitter} framework. The HERA $e^{\pm} p$ data are used first by relaxing some sea parameters in the HERAPDF2.0 with unsuppressed strangeness. The momentum fraction difference of anti-down and anti-up quark density, $x(\dbar -\ubar)$, is found to be below zero towards larger momentum fractions, namely at $x\in (10^{-2}, 1)$, indicating that proton carries more anti-up quarks than anti-down quarks. 
But because the $e^{\pm} p$ DIS cross section can not constrain $\dbar$ and $\sbar$ separately, while it has a strong sensitivity to $\ubar$ and $\dbar$, it is hard to determine $\dbar$ distribution with exact shape. To solve this problem, we include, in our second fit, the ATLAS $W^{\pm}/Z$ data which can impact on all of the sea, valence quarks and gluon distributions. After the ATLAS $W^{\pm}/Z$ data are included, we find that not only the individual parton flavors are well determined, but also the corresponding uncertainties are reduced. The obtained results on $S_G$ in two $x$ regions, $0.004 \leq x \leq 0.8$ and $0 \leq x \leq 1$ at $Q^2= 4~\mathrm{GeV}^2$, and $x(\dbar -\ubar)$, $\dbar/\ubar$ from both the HERAshape and the ATLASshape PDFs indicate that there is an excess of anti-up quark over anti-down quark in the momentum fraction range $x\in(10^{-2}, 1)$. The other worldwide PDFs also support this idea within their uncertainties. As an improved results of the HERAshape PDFs,
the results obtained from alternative parameterization in the ATLASshape PDF provides more solid basis to reach the main standpoint of this paper.

\section{ACKNOWLEDGMENTS}
We thank the \texttt{xFitter} developers for providing useful help during this work and some online discussions on the usage of \texttt{xFtter} package and other programs mentioned in this paper.
This work is supported by National Natural Science Foundation of China (Grants No.~12335006 and No.~12075003). This work is also supported by High-performance Computing Platform of Peking University.

\bibliographystyle{elsarticle-num}

\end{document}